\documentclass[doublecol]{epl2} 

\title{Universal correlators \& distributions as experimental signatures of 2+1 dimensional Kardar-Parisi-Zhang growth}

\author{Timothy Halpin-Healy\inst{1} and George Palasantzas\inst{2}}

\institute{
\inst{1} Physics Department, Barnard College, Columbia University, New York NY 10027\\
\inst{2} Department of Applied Physics, University of Groningen, Nijenborgh 4, 9747 AG Groningen, The Netherlands
}

\abstract{We examine height-height correlations in the transient growth regime of  the 2+1 Kardar-Parisi-Zhang (KPZ) universality class, with a particular focus on the {\it spatial covariance} of the underlying two-point statistics, higher-dimensional analog of the 1+1 KPZ Class Airy$_1$ process.  Making comparison to AFM kinetic roughening data in 2d organic thin films, we use our universal 2+1 KPZ spatial covariance to extract key scaling parameters for this experimental system.  Additionally, we explore the i) height, ii) local roughness, and iii) extreme value distributions characteristic of these oligomer films, finding compelling agreement in all instances with our numerical integration of the KPZ equation itself.  Finally, investigating nonequilibrium relaxation phenomena exhibited by 2+1 KPZ Class models, we have unearthed a universal KPZ ageing kinetics. In experiments with ample data in the time domain, our 2+1 KPZ Euler {\it temporal covariance} will allow a quick, independent estimate of the central KPZ scaling parameter.
 }

\begin{document}

\maketitle
\section{Introduction}
Recent years have witnessed spectacular advances~\cite{KK}, on both experimental and theoretical  fronts, in the nonequilibrium statistical mechanics of the Kardar-Parisi-Zhang (KPZ) equation in 1+1 dimensions.  On the experimental side, Takeuchi and coworkers~\cite{KT} have verified in extraordinary detail that the stochastic, interfacial fluctuations in turbulent liquid crystals are governed by the Tracy-Widom (TW) limit distributions~\cite{TW}, known long ago to Pr\"ahofer and Spohn~\cite{PS00}, as well as Johannson~\cite{KJ}, in their seminal studies of the polynuclear \& single-step growth models, canonical members of the KPZ universality class. Beyond the universal TW height fluctuations, these experiments also provided strong evidence for the underlying 2-pt spatial correlations, known in the flat geometry to be set by the covariance of the Airy$_1$ process~\cite{PS04}.  Impressive theoretical efforts emerged first on curved~\cite{SS}, then flat KPZ problems, the latter due to Calabrese and Le Doussal~\cite{Cala}. Shortly thereafter, Imamura \& Sasamoto~\cite{IS}, in a technical tour-de-force, extracted the {\it stationary-state} statistics characteristic of the 1+1 KPZ Class, dictated by the Baik-Rains limit distribution~\cite{BR}. Subsequent work on the KPZ stationary-state by Takeuchi has established a precursor experimental signature~\cite{KT13}, providing access to dynamics in this elusive KPZ regime.  Numerical studies~\cite{HH14,Braz1+1,BrazEPL}, invoking KPZ scaling theory~\cite{KMHH}, built upon the Krug-Meakin (KM) toolbox~\cite{KM}, and inspired by early efforts~\cite{KBM,HH91} to distill essential KPZ distributions, have done much to bolster our understanding of universality within the 1+1 KPZ Class; they inform the experimental findings and, furthermore, complement the powerful, but often quite model-specific mathematical developments.

Here, we step up to the 2+1 KPZ Class, addressing the business of higher-dimensional stochastic growth.   We begin with the Kardar-Parisi-Zhang equation~\cite{KPZ},  which characterizes the fluctuations of the height $h({\bf x},t)$ of a kinetically roughened interface:
$$\partial_th=\nu\nabla^2 h +{1\over 2}\lambda(\nabla h)^2 + \surd D \eta,$$
\noindent in which $\nu,\hspace{0.5mm}\lambda\hspace{0.5mm}$ and $D$ are phenomenological parameters, the last setting the strength of the stochastic noise $\eta.$   
For the flat KPZ Class geometry in a system size of lateral dimension $L,$ the width $w$ of the fluctuating interface grows with time t as $w(t,L)$=$\surd\langle h^2\rangle-\langle h\rangle^2\sim t^{\beta}F(L/t^{\beta/\chi})$, where $\beta$, $\chi$ and $z$=$\chi/\beta$ are, resp., the early-time roughness, saturation width, and dynamic exponents, and $F$ the universal Family-Vicsek scaling function~\cite{FV}.  Halpin-Healy~\cite{HH12,HH13} has made a large-scale numerical integration of the 2+1 KPZ equation, deep in the nonlinear regime ($\lambda$=20), with $L$=10$^4$ and exhaustive averaging now, with this work, reaching $\sim$10$^3$ runs.  This 2+1 KPZ Euler investigation was complemented by a quartet of zero-temperature transfer matrix studies of directed polymers in random media (DPRM), as well as kinetic roughening simulations of 2+1 ``restricted-solid-on-solid" (RSOS)  and 2d driven dimer models.  In a first pass, the traditional {\it zero-mean, unit-variance} height distribution (HD) was obtained, revealing a skewness $s_{2+1}\hspace{-1.0mm}\approx$ 0.424 and kurtosis $k_{2+1}\hspace{-1.0mm}\approx$ 0.35; see, too~\cite{BrazRC}.  Via convincing data collapse of all 7 studied models, the case was thus made for 2+1 KPZ Class universality~\cite{HH12}.   With much greater effort,  and relying upon full use of the KM toolbox to strip numerical datasets of model-dependent baggage, the underlying, universal limit distribution was isolated; i.e., the fundamental PDF at the very heart of 2+1 KPZ Class universality~\cite{HH12,HH13}.

We recall in 2+1, there are no exact results, aside from the sacred dimension-{\it independent} KPZ exponent identity: $\chi$+z=2. Stretching back 25 years, there has been intense activity within the KPZ community to pin down both $\beta$ \& $\chi$.  Most recently, a multi-model examination~\cite{HH13} of temporal correlations in RSOS, KPZ Euler, and Gaussian DPRM stationary-states, has directly estimated $\beta_{2+1}$=0.241$\pm$0.001, in agreement with  the landmark effort of Forrest, Tang \& Wolf~\cite{FTW}, who found 0.240$\pm$0.001 for a hypercubic-stacking model in the transient growth regime. Independent, state-of-the-art estimates for the saturation width exponent, relying upon clever code implementations and two quite closely related discrete models-RSOS~\cite{Parisi} and driven-dimers~\cite{KO}, have produced the result $\chi_{2+1}$=0.393$\pm$0.003, though the $\beta$ findings, aided by the KPZ identity, would suggest a value nearer to 0.387-0.389.  While the numerical assault continues, the essential 2+1 KPZ portrait~\cite{HH12,HH13} is now solidly in place, awaiting experimental tests and, ultimately, an exact solution; i.e., higher dimensional analogs of the 1+1 KPZ Class TW-GOE limit distribution and Airy$_1$ process.

\section{2+1 KPZ Class: Height Fluctuations} 
\noindent Despite a flurry of experimental works in the early years studying potential 2+1 KPZ systems, such as etching, MBE, and chemical vapor
deposition~\cite{Krug}, it was discovered that surface diffusion phenomena, coupled to large-scale morphological instabilities such as mounding and grain boundaries, often scuttled 
KPZ hopes. Towards the end of this period, however, Palasantzas and coworkers~\cite{GP}, studying the growth of 
vapor-deposited organic thin films, extracted exponents, $\beta$=0.28$\pm$0.05 and $\chi$=0.45$\pm$0.04, highly suggestive
of accepted KPZ values, but most importantly, revealed a height histogram manifesting a ``pronounced" right tail, characterized by a distinctly {\it positive skewness.}  With a deeper understanding of 2+1 KPZ universality~\cite{HH12,HH13} and noting new, related work on semiconductor films~\cite{Almeida}, we have returned to our raw experimental data-sets to ascertain what might be understood of that long, ``pronounced" tail. In fact, we have discovered much richness therein.  While referring the reader to the original paper for full details~\cite{GP}, we mention specifically that: i) the five-ring poly-[p-phenyl-vinylene] oligomer Ooct-OPV5,
shown within Figure 1, has a characteristic length of 2.9nm, was ii) vapor-deposited upon a well-prepared Si substrate held at room temperature (RT), iii) the resulting kinetically roughened thin organic film growing to an eventual thickness of 300nm, at an average rate $v_\infty\hspace{-2mm}\approx$7.1nm/min.
The film surface morphology was captured by an AFM (Digital Instrument Nanoscope IIIa) in tapping mode, using a 512x512 pixel array to render seven images of linear size 2$\mu$m.  See Fig 1 inset for a representative false color AFM snapshot; note scale bar, indicating the vertical range of surface roughness, as well as the Ooct-OPV5 chemical schematic and molecular model.  

\begin{figure}
\includegraphics[angle=0,width=85mm]{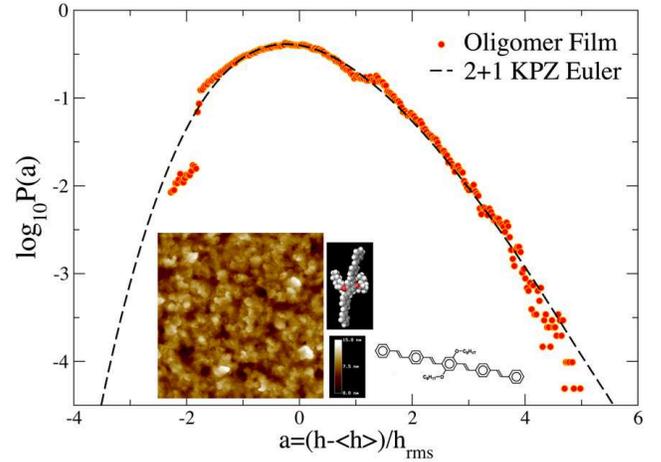}
\caption{\label{fig:Fig1}Height Fluctuation PDFs: 2+1 KPZ Equation vs. kinetic roughening experiment. Inset: AFM image- 50nm thick oligomer film; lateral scan dimension 2$\mu$m.}
\end{figure}

Averaging over these AFM thin film scans~\cite{GP}, we craft here the traditional zero-mean, unit-variance HD and, in Figure 1 proper, make comparison to the height distribution of our 2+1 KPZ Euler integration~\cite{HH12}. It is clear that the ``pronounced" right tail, as fully revealed in our semilog plot, is {\it distinctly KPZ in character;} indeed, well down to probabilities of order 10$^{-4}$, thus confirming this system's long-deserved membership within the 2+1 KPZ Class. 
We believe Ooct-OPV's linear shape generates a ballistic deposit $(\lambda\textgreater 0)$ of short, oriented rods, a notion supported by the HD's skewness, which correlates with the sign of the KPZ nonlinearity. 
The small kink at $a\hspace{-1mm}\approx\hspace{-1mm} 1^+$ arises because, at this film thickness, $h_{rms}\hspace{-1mm}\approx$1.8nm, indicating a slight {\it excess} for height jumps $\Delta h\hspace{-1mm}\approx$2-3nm, precisely that expected for a predominantly vertical, but tilted oligomer molecule. We recall that 1.7nm terrace steps were observed in lamellae-type grains in related pentacene deposits~\cite{knuckles}.  Finally, we suggest $\pi$-``stacking" to underlie the sticking mechanism implicit to our ballistic deposition interpretation. Here- the embracing, hydrocarbon arms at the midriff of the aromatic phenyl-backbone would enhance this essential {\it intermolecular} noncovalent bonding process but, of course, would likely spoil the alternating, tilted-herringbone motif typically seen in layer-by-layer growth of pentacene deposits; see Ref~\cite{knuckles}, Fig.4. 

\section{2+1 KPZ Class: Spatial Covariance} 
Historically, the kinetic roughening correlators of interest have been:
i) the equal-time {\it height-difference} correlation function:
$$C_{h}(r)\equiv\langle (h({\bf x}+{\bf r},t)-h({\bf x},t))^2\rangle=A_{h}r^{2\chi}g(s/t^z)$$
and its near relative, the 
ii) KPZ {\it spatial covariance-}
$$C_v(r)\equiv\langle h({\bf x}+{\bf r},t)h({\bf x},t)\rangle-\langle h\rangle^2$$
so that $C_{h}=2w^2 - 2C_v$.   Additional, important ingredients include-  
iii) the KPZ finite-size scaling result of Krug \& Meakin~\cite{KM}, which predicts a small shift in the asymptotic growth velocity: 
$$\Delta v=v_L-v_\infty= - \frac{1}{2}\lambda A/L^{2-2\chi}$$
in a system of size $L$, as well as-
iv) the characteristic, tilt-dependent KPZ growth velocity $v$=$v_\infty$+$\frac{1}{2}\lambda(\frac{\delta h}{\delta x})^2$, quadratic function of the angle of the canted substrate.  The latter relations permit determination of the key model-dependent parameters $v_\infty$ and 
$\theta$=$A^{1/\chi}\lambda$, the essential quantities of KPZ scaling theory~\cite{KMHH}.  
Heavy use of this KM toolbox was necessary to numerically extract the universal 2+1 KPZ Class limit distribution proper~\cite{HH12}.

Here, in complementary work, we distill the {\it universal} 2+1 KPZ spatial covariance, higher-dimensional analog of Airy$_1$ process.  To realize this, we must fix the connection between the KM parameter $A$ and the prefactor, $A_{h},$ in the height difference correlation function.  In 1+1, $A_h$=$A$, but for 2+1 KPZ, the Fourier integrals conspire to break this equivalence~\cite{HH13}; consideration of a unit-amplitude momentum-space correlator, $\langle\vert h_k\vert^2 \rangle\equiv L^2k^{-(2+2\chi)},$ yields a 2+1 Krug-Meakin formula- $\Delta v(L)=\frac{1}{2}\lambda I_{KM}$,  with the 2d $k$-space integral:
$$I_{KM}=\int_0^{\frac{\pi}{L}}\frac{dk_x}{2\pi}\int_0^{\frac{\pi}{L}}\frac{dk_y}{2\pi}k^2\langle\vert h_k\vert^2 \rangle$$
Calculated in rectangular coordinates appropriate to our numerical simulation in an $L$x$L$ box, this leads to an intriguing integral over the unit square-
$$A\equiv I_{KM}/L^{2-2\chi}=\pi^{-2\chi}\int_0^1d\eta\int_0^1d\omega/(\eta^2+\omega^2)^{\chi}$$
which, for $\chi_{2+1}$=0.390, is numerically evaluated to yield $A$= 0.6077.  The same unit-amplitude $k-$space correlator, when Fourier-transformed alone, produces the value  
$A_{h}$= $-\Gamma(-0.39)/2^{1.78}\pi\Gamma(1.39)$=0.3926 for the height-difference correlation function prefactor.  Hence, $A_{h}/A$=0.6460,
allowing us to extract, from our measured~\cite{HH12}, system-specific KM parameter $A$, the corresponding value of $A_{h}$ needed, now, to generate a plot of the 2+1 KPZ spatial covariance. 

\begin{figure}
\includegraphics[angle=0,width=85mm]{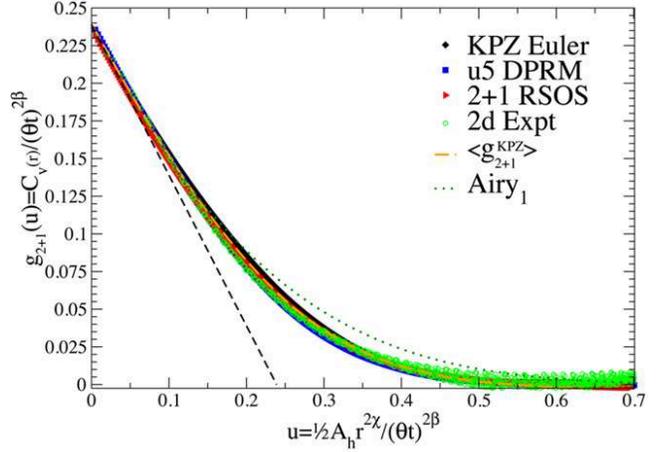}
\caption{\label{fig:Fig1} 2+1 KPZ Class: Universal Spatial Covariance.}
\end{figure}

Thus, we show in Fig. 2, our 2+1 KPZ Euler results with $L$=10$^4$, evolved to a time t=2k, with nonlinearity $\lambda$=20. From our KM analysis\cite{HH12} of the 2+1 KPZ equation, we know $A$=0.02295; hence, correlator prefactor $A_{h}$=0.01483. As is custom from 1+1 KPZ Class Airy$_1$ discussions, both horizontal and vertical axes have been scaled by the factor $(\theta t)^{2\beta}$, with $\theta$=1.192x10$^{-3}$ for 2+1 KPZ Euler.   Note, firstly, that the vertical intercept, where $r$=0, corresponds to the universal 2+1 KPZ Class variance, $\langle\xi^2\rangle_c$=$w^2/(\theta t)^{2\beta}$, which we estimate here as 0.237, extrapolating a linear fit to the axis. This agrees quite nicely with our independent 2+1 KPZ Euler estimate, 0.243, obtained previously, as well as the value 0.235, representing the mean there over 7 distinct KPZ Class models~\cite{HH12}. Secondly, the scaled abscissa can be understood as $u\hspace{-1mm}\sim\hspace{-1mm}A_{h}(r/\xi_\parallel)^{2\chi}$
reminding us that the parallel correlation length
$\xi_\parallel$ sets the key scale for all matters of growth upon the surface. Indeed, this axis rescaling demands that the covariance have unit slope at its vertical intercept; see Fig. 2, dashed line.  We include, as well, large-scale 2+1 RSOS \& DPRM results, for which
we have~\cite{HH12}:  $\theta$=0.66144 \& 0.2518 and $A_{h}$=0.7755 \& 0.7738, respectively. We used $\beta$=0.241 \& $\chi$=0.390 for all the three models, the good data collapse evidence of a {\it universal} 2+1 KPZ Class spatial covariance, labeled $\langle g_{2+1}^{KPZ}\rangle$ in the plot, an average of KPZ Euler, RSOS and DPRM results. While AFM scans from thin film experiments have traditionally been used to investigate the 2-pt height difference correlation function, here we analyze the oligomer data sets making comparison to our universal 2+1 KPZ spatial covariance.
This involves a two parameter fit- first, the vertical intercept of the {\it nonuniversal} experimental trace (not shown) is scaled to match the 2+1 KPZ Class variance $\langle\xi^2\rangle_c$=0.237. Numerically, this scale factor, as per discussion above, determines the quantity $(\theta t)^{\beta}$=3.71.  Since the growth rate, 7.1nm/min, is known and the AFM scans were taken for a 50nm thick film, the elapsed time $t$=422sec is then fixed, which allows us to estimate the key KPZ scaling parameter $\theta_{expt}\hspace{-1.5mm}\approx$0.256 for this experimental system. Next, rescaling the abscissa of the data to optimize the fit to our universal 2+1 KPZ spatial covariance, we determine $A_h\hspace{-1mm}\approx$0.165, which implies Krug-Meakin $A_{expt}\hspace{-1mm}\approx$0.255. Knowing $A$ \& $\theta,$ we thereby estimate $\lambda_{expt}$=5.5$\pm$0.5 nm/s for our kinetically roughened RT oligomer films, a not unreasonable value given the geometric particularities of the deposited molecules.  With our universal 2+1 KPZ Euler spatial covariance, we therefore manage a successful determination of the KPZ nonlinearity $\lambda$ in an actual 2d experiment.
    
\section{2+1 KPZ Class: Local Roughness} 

Beyond the transient-regime $(\xi_\parallel\hspace{-1mm}\ll\hspace{-1mm}L)$ HDs, there has been long-standing interest
in the steady-state $(\xi_\parallel\hspace{-1mm}\gg\hspace{-1mm}L)$ {\it width distribution} of such fluctuating interfaces. The crucial work of R\'acz and Plischke~\cite{MP}, on these and related matters, has greatly informed our own efforts here.  We encourage the dedicated reader to have a look at the experimentally-motivated, penultimate  section of this well-written, classic paper.
The relevant variable is now the squared width: $w^2=\langle h^2\rangle-\langle h\rangle^2$.
Implicit in the angular brackets is the idea of spatial averaging over a system of linear dimension
L under the assumption of periodic boundary conditions (PBC), an ensemble average over many realizations
of the noise, as well as extraction of the large-L asymptotics.  Early numerical efforts~\cite{Foltin} on such steady-state 1d KPZ {\it random-walk} interfaces
indicated that the underlying ``squared-roughness" distribution $P(w^2)$ possessed a purely {\it exponential} right tail, a result suggested by an approximate analytical treatment, generalized in later work~\cite{Antal}.  The 2+1 KPZ situation remains open, with numerous works weighing in on the matter of a ``stretched" exponential tail in this dimension~\cite{Parisi,FDA,KO}. It should be stressed that for these numerical investigations, the goals demand pushing the system, for fixed large L and PBC, deep into the steady-state to extract the universal, scaled distribution $P(w^2/\langle w^2\rangle)$. Unfortunately, PBC are a luxury not permitted the KPZ experimentalist.  The real-world necessities in that realm preclude PBC, dictating instead window boundary conditions- WBC. 

\begin{figure}
\includegraphics[angle=0,width=85mm]{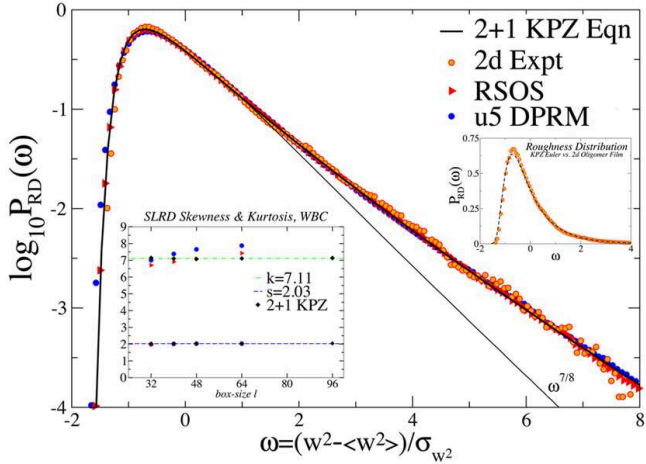}
\caption{\label{fig:Fig1} 2+1 KPZ Class: Local Roughness Distribution.}
\end{figure}

To this end, motivated by classic theory~\cite{MP}, as well as recent experimental work on semiconductor films~\cite{Almeida}, we calculate here the squared, {\it local} roughness distribution (SLRD) for the 2+1 KPZ equation itself, using various values of the WBC ``box"-size $\ell.$ In Figure 3, we show our high-precision 2+1 KPZ Euler SLRD results, complemented by DPRM \& RSOS findings, along with organic 2d film experimental data~\cite{GP}, all with box-size $\ell$=40.  The agreement is excellent and quite forthcoming, revealing the SLRD to be a salient experimental signature of 2+1 KPZ kinetic roughening.  We measure for the 2+1 KPZ Euler SLRD, with WBC, a skewness $s$=2.03 and kurtosis $k$=7.11, obtained via exhaustive averaging over our huge numerical data sets.  These KPZ values, see Fig 3 inset, are {\it independent} of box-size for $\ell\in$(32,96).  That is, for $\ell\hspace{-1.5mm}\ll\hspace{-1.5mm}\xi_\parallel^{KPZ}$, our 2+1 KPZ Euler statistics {\it are stationary,}  yielding constant quantities.  We emphasize that there is no interest at all here in the limit of large $\ell$; quite the contrary, one needs small window sizes experimentally.  These are thin films, so $\xi_\parallel$ remains small.  In any case, our organic film data yields, for $\ell$=40, $(s,k)$=(2.12,7.77); related 2+1 KPZ Class models- DPRM: $(s,k)$=(2.01,7.38), RSOS: $(s,k)$=(2.00,6.91). The WBC SLRD of the latter was the subject of a quite preliminary, though helpful study~\cite{Paiva}, reiterating the distinction between PBC \& WBC~\cite{MP,Antal}.  One needs to take care, however, since for these models, the correlation length, $\xi_\parallel\hspace{-1.5mm}\sim\hspace{-1.5mm}(\surd A\lambda t)^{1/z}$, grows much more slowly than KPZ with $\lambda$=20. In our work here, we have greatly privileged 2+1 KPZ Euler, for which the key combination $\lambda\surd A$=3.03, so $\xi_\parallel\hspace{-1.5mm}\sim\hspace{-1.0mm}200$ at $t$=2k, considerably larger than that of RSOS (\& DPRM) for comparable simulation times~\cite{HH13}.
Finally, plotting $\omega^{7/8}$ as the abscissa of our 2+1 KPZ Euler SLRD data set, rids the tail of its ``stretch"; the resulting thin black trace, our final addition to Figure 3, runs true to any straight-edge laid upon it.

\section{2+1 KPZ Class: Extremal Statistics} 
Within the KPZ kinetic roughening context, a third relevant  distribution captures the extreme-value (EV) statistics of the height fluctuations, providing a well-defined arena in which to examine extremal behaviors of {\it correlated} random variables.  Here, inaugural numerics for the 1+1 KPZ Class by Shapir \& collaborators~\cite{SRay},  were quickly followed up by important analytical  
works~\cite{Satya123}. Interestingly, the 2+1 KPZ case, with WBC appropriate to 2d experiments, remains untouched.  Again, we highlight our 2+1 KPZ Euler integration, though we have results, too, for DPRM \& RSOS models. In Figure 4, we make explicit comparison to oligomer data sets, with box-size $\ell$=40$\ll\hspace{-1.5mm}\xi_\parallel^{KPZ},$ assuming WBC. Note we have plotted both the Maximal and Minimal {\it relative height distributions} (MRHD) for the experimental data, but only a single 2+1 KPZ Euler trace. This, because our KPZ numerics indicate no experimentally relevant distinction between the two {\it scaled} MRHDs; that is, for the KPZ Max/Min relative height deviation, we measure skewness $s$=0.884/0.877  and kurtosis $k$=1.20/1.172.  The corresponding values for our 2+1 DPRM: $s$=0.845/0.860 and $k$=1.196/1.151, respectively.  Thus, the difference between Max \& Min PDFs for a given model are {\it distinctly smaller} than the differences {\it between} models for a given distribution.  We see identical MRHDs in our RSOS work too.  Our WBC findings stand in sharp contrast to previous 2+1 KPZ growth model studies~\cite{T08} with small $L$ and PBC, which indicated $s_{max}\hspace{-1.0mm}\approx$0.79 \& $s_{min}\hspace{-1.0mm}\approx$0.65, results we have confirmed, finding 0.818 \& 0.649.  We emphasize again, however, that such PBC values are not germane to the experimental matter at hand; it is the newly reported WBC results that are relevant and, thus, have been recorded in Figure 4 and underlie the high precision 2+1 KPZ Euler trace we compare directly to experiment.

\begin{figure}
\includegraphics[angle=0,width=87mm]{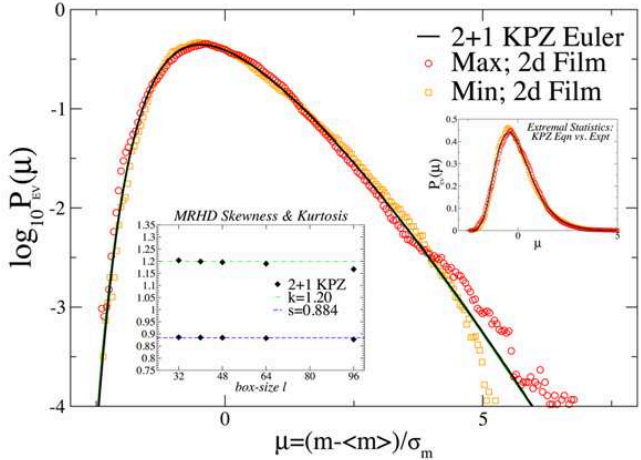}
\caption{\label{fig:Fig1} 2+1 KPZ Class: Extremal Height Fluctuations.}
\end{figure}

\section{2+1 KPZ Persistence: Temporal Covariance}
In a rather separate, but intriguing line of inquiry, work on the nonequilibrium {\it relaxation properties} of KPZ surface growth~\cite{Krech}, has revealed simple ``ageing" behavior~\cite{KK97,Malted,GO14}, without the need for additional exponents; that is, the universality classes are dictated entirely by the dynamical exponent $z$. In fact, the inevitable, but very natural conclusion following from these studies is that the full machinery of KPZ scaling theory~\cite{KMHH,HH12}, recently applied with great success to the 2+1 KPZ Class kinetic roughening phenomena, should be applicable to KPZ ageing kinetics as well; indeed, we show precisely this in what follows. The entry-level object of interest is the {\it two-time} spatiotemporal correlator for the KPZ height variable~\cite{Krech,KK97,Malted}:
$$C(t,s;{\bf r})=\langle h(t,{\bf r})h(s,{\bf 0})\rangle -\langle h(t)\rangle\langle h(s)\rangle$$
\noindent For $t$=$s$, of course, we retrieve the spatial correlator discussed earlier; hence, here, an extended Family-Vicsek scaling
prevails-
$$C(t,s;{\bf r})=s^{2\beta}F_C(\frac{t}{s},\frac{\mid\hspace{-1.0mm}{\bf r}\hspace{-1.0mm}\mid^z}{s})$$
with $F_C$ a  new universal scaling function. Our specific focus now, however, is the temporal autocorrelator, so we set {\bf r}=0 and consider
ageing intervals $t$-$s$, much greater than the initial {\it waiting time} $s$ needed to generate a sufficiently mature KPZ interface.
We have investigated this possibility for our trio of 2+1 KPZ Class models, using system sizes $L$=10$^4$ and making many runs in all instances, with $s$=100 for RSOS \& DPRM simulations, and $s$=5 for KPZ Euler.  In crafting our Figure 5 temporal covariance, we have achieved , as was the case for Figure 2 spatial covariance, complete data collapse by scaling the ``time", here it is the {\it waiting time} $s$, by the model-dependent parameter $\theta$=$A^{1/\chi}\lambda$ for each of the individual data sets; i.e., $\theta$=1.192x10$^{-3}$, 0.2518, 0.66144 for 2+1 KPZ, DPRM, RSOS, respectively~\cite{HH12}.  In this manner, one sees clearly the emergence of a universal 2+1 KPZ {\it temporal covariance,} which convincingly captures the essence of ageing kinetics in this context. The dashed trace,  calculated as an equally weighted average of our three model simulations, represents our best determination of this universal 2+1 KPZ autocorrelator.  Additionally, our rough estimate for the autocorrelation exponent, -1.17(2), reflects a simple linear fit, for $t/s\textgreater4,$ to the power-law, $F_C(t/s,0)=C(t,s;{\bf 0})/s^{0.483}\sim (t/s)^{-1.17}$, evident in our double-log plot, assuming the Kelling-Odor value~\cite{KO} $\beta_{2+1}$=0.2415. 
Independent, large scale numerics~\cite{GO14} on 2d Driven Dimers, a planar lattice gas generalizing the 1+1 KPZ Class single-step model to higher dimension, fits nicely into this picture of universal 2+1 KPZ ageing, and yields an exponent -1.2.  These values are not far at all from the Kallabis-Krug conjecture~\cite{KK97}, $-d/z\hspace{-1.0mm}\approx$ -1.24, in this dimension.
Further numerical aspects, including extension of this richer universality to the 2+1 KPZ {\it autoresponse} function, will be discussed elsewhere. 

Our immediate goal regarding these matters is to propose the universal 2+1 KPZ temporal correlator as a benchmark against which experimentalists can measure the model-dependent KPZ scaling parameter $\theta$ appropriate to their system.  This may not be feasible for 2d thin films, where the richness of the data lies in its spatial, not temporal, aspects.  However, there may come to pass experimental realizations, here- the turbulent liquid-crystal work~\cite{KT} comes to mind, in which there is a wealth of information available in the time domain.  Should this be the case, matching the experimental autocorrelator to our {\it universal} 2+1 KPZ {\it temporal covariance} permits an independent
assessment of $\theta_{expt}$, necessary to go beyond the simple HD and gain access to an experimental portrait of the underlying, fundamental 2+1 KPZ Class limit distribution, higher dimensional analog of 1+1 KPZ Tracy-Widom~\cite{KK,TW}. Our numerics indicate that 10$^8$ data points (i.e., ten 10-megapixel images) would be sufficient to get the job done.
Of course, more is better, but the slope and location of the temporal 2+1 KPZ autocorrelator emerges quite readily in our experience;
the essential purpose of massive ensemble averaging merely to quell the noise and pin down the exponent.

\begin{figure} 
\includegraphics[angle=0,width=85mm]{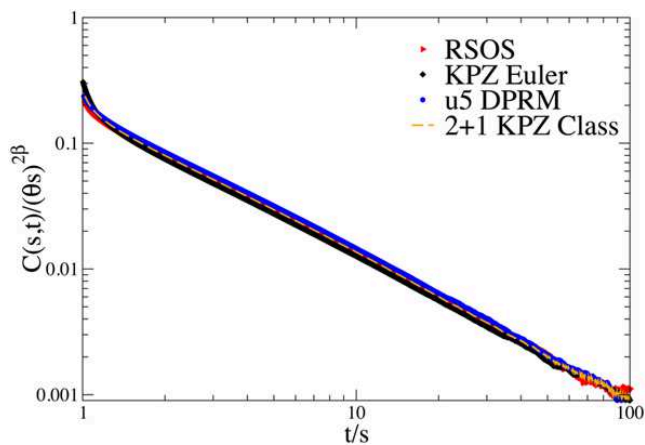}
\caption{\label{fig:Fig} 2+1 KPZ Class: Universal Temporal Covariance.}
\end{figure}

{\bf Summary. --}   
Using the wisdom of KPZ scaling theory, we have successfully distilled the universal spatial \& temporal covariances characteristic of the 2+1 KPZ Class.  In the former instance, we have revisited kinetic roughening data of organic thin films~\cite{GP}, extracting the KPZ nonlinearity $\lambda$; the latter case awaits an experimental application of our universal KPZ ageing kinetics. Additionally, with our 2+1 KPZ Euler integration, we have made high-precision, numerical investigations of the squared local roughness (SLRD) and extremal, relative height (MRHD) distributions, complementing prior work on the 2+1 KPZ class HD~\cite {HH12}.  Our analysis of oligomer Ooct-OPV5/Si data sets reveals it to be a system solidly within the 2+1 KPZ universality class. 
Armed with 3 distributions, 2 universal correlators, and an amply versatile Krug-Meakin toolbox~\cite{KMHH}, one sees the beginning of a new era in higher-dimensional KPZ experimental work.

\acknowledgments
THH would like to express his gratitude to Wim van Saarloos for a generous sabbatical visit to the Netherlands that triggered the present collaboration, John Magyar for Schr\"odinger MD simulations of Ooct-OPV5, G\'eza \'Odor concerning KPZ ageing, and F. Bornemann for his 1+1 Airy$_1$ trace, which we rescaled for comparative purpose.

\end{document}